\documentclass[12pt]{article}
\usepackage[margin=3cm]{geometry}
\usepackage{amsfonts}
\usepackage{amsmath}
\usepackage{amssymb}
\usepackage{helvet}
\usepackage{enumerate}
\usepackage{framed}   
\usepackage{graphicx}
\usepackage{subfigure}
\usepackage{verbatim}
\usepackage[bookmarks=false,colorlinks=true]{hyperref}
\usepackage{color,soul} 
\usepackage{url}
\usepackage[normalem]{ulem} 

\newtheorem{thm}{Theorem}

\newtheorem{cor}{Corollary}

\newtheorem{prf}{Proof}

\newtheorem{rem}{Remark}
\newtheorem{ex}{Example}

\def\figs{Figures}

\newcommand\qed{\ensuremath{\	\blacksquare}}	

\title{A Note on Parallel Algorithmic Speedup Bounds}
\author{Neil J. Gunther} 
\date{February 8, 2011}

\begin{document}
\maketitle

\begin{abstract}
A parallel program can be represented as a directed acyclic graph.
An important performance bound is the time $T_\infty$ to execute the critical path 
through the graph. We show how this performance metric is related to Amdahl speedup and 
the degree of average parallelism. These bounds formally exclude superlinear performance.
\end{abstract}

\section{Computational DAG}
\begin{figure}[!hb]
\centering
\includegraphics[scale = 0.5]{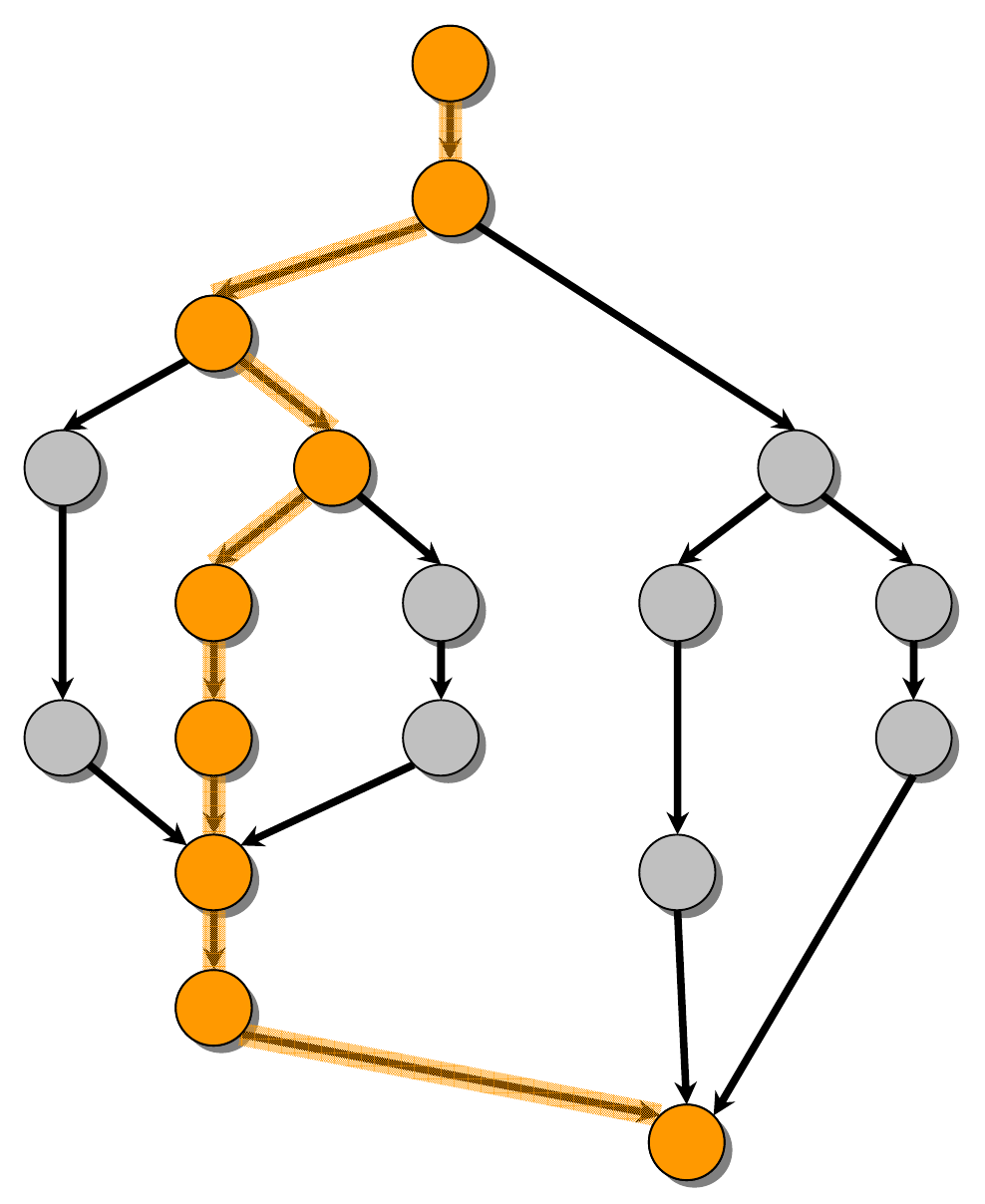}  
\caption{Critical path ({\em orange}) in an parallel task graph~\cite{leiserson}} 
\label{fig:dag}
\end{figure}

A parallel program can be represented as a directed acyclic graph (DAG),
where nodes correspond to tasks (or subtasks) and arrows represent control
or communication between tasks. Leiserson~\cite{leiserson} characterizes
the performance of parallel programs by the elapsed time $T_1$ to execute
all the nodes in a DAG (e.g., Fig.~\ref{fig:dag}), and the time $T_\infty$
to execute the {\em critical path}.

In project management, a critical path is the sequence of project network
activities (e.g., a PERT chart) which add up to the {\em longest} overall duration. It
determines the shortest possible time to complete the project. Any delay of
an activity on the critical path directly impacts the planned project
completion date (i.e. there is no float on the critical path). A project
can have more than one critical path.

The time to execute a program on $p$ processors is $T_p$ and the speedup metric is:
\begin{equation}
S_p = \dfrac{T_1}{T_p} \label{eqn:speedup}
\end{equation}
with computational efficiency:
\begin{equation}
E_p = \dfrac{S_p}{p}   \label{eqn:efficiency}
\end{equation}
i.e., the average amount of speedup per processor.

\section{Performance Bounds} \label{sec:bounds}
Leiserson~\cite{leiserson} claims there are two {\em lower} bounds on 
parallel performance for Fig.~\ref{fig:dag}:
\begin{align}
T_p &\geq T_1/p 	\label{eqn:linear} \\
T_p &\geq T_\infty 	\label{eqn:ceiling}
\end{align}
$T_1/p$ is the reduced execution time attained by partitioning the work 
(equally) across $p$ processors. 
Clearly, $T_p$ cannot be less than the time it takes to
execute a $p$-th of the work---the meaning of (\ref{eqn:linear}). Similarly, $T_p$ cannot be less than the time
it takes to execute the critical path, even if there are an infinite number
of physical processors---the meaning of (\ref{eqn:ceiling}).

Substituting (\ref{eqn:linear}) into (\ref{eqn:speedup}):
\begin{equation}
S_p = \dfrac{T_1}{T_1/p} = p \label{eqn:linspeed}
\end{equation}
which corresponds to ideal {\em linear} speedup.
In reality, we expect the speedup to be generally sublinear:
\begin{equation}
S_p \leq p
\end{equation}
Under certain special circumstances speedup may exhibit {\em superlinear} performance:
\begin{equation}
S_p > p \label{eqn:superlin}
\end{equation}
Leiserson excludes (\ref{eqn:superlin}) on the basis of (\ref{eqn:linear}). 
He also states that because of (\ref{eqn:ceiling}), the {\em maximum} 
possible speedup is given by:
\begin{equation}
S_\infty = \dfrac{T_1}{T_\infty} \label{eqn:parallelism}
\end{equation}
He calls (\ref{eqn:parallelism}) the ``parallelism'' and it 
corresponds to the average amount of work-per-node along the critical path.
But what do these bounds really mean?

\subsection{Example}
Consider an example based on Fig.~\ref{fig:dag}.

\begin{ex}
Following Leiserson, let's assume for simplicity that each node in the DAG takes just 
1 unit of time to execute. Then, the total time to execute the entire DAG 
on a single processor is $T_1 = 18$ time steps.

Similarly, the \uline{critical path} contains nine nodes, so $T_\infty = 9$ and 
from (\ref{eqn:parallelism}):
\begin{equation}
S_\infty = \dfrac{18}{9} = 2 \label{eqn:para2}
\end{equation}
Hence, the \uline{maximum} possible speedup is 2.
\end{ex}
Note, however, that this maximum speedup (\ref{eqn:parallelism}) is not the same as the more familiar 
Amdahl bound~\cite{ppa,geo,usl}:
\begin{equation}
S_\infty^{\rm Amdahl} = \dfrac{1}{\sigma} \label{eqn:amdasymp}
\end{equation}

Equation (\ref{eqn:amdasymp}) is the asymptotic form of the Amdahl speedup function~\cite{ppa}:
\begin{equation}
S_p^{\rm Amdahl} = \dfrac{p}{1 + \sigma (p-1)} \label{eqn:Amdahl}
\end{equation}
in the limit of an infinite number of processors $p \rightarrow \infty$.
In Fig.~\ref{fig:dag}, the {\em serial fraction} ($\sigma$) corresponds to 4
single nodes out of 18 total nodes and therefore:
\begin{equation}
S_\infty^{\rm Amdahl} = \dfrac{18}{4} = 4.5
\end{equation}
which is numerically greater than the ``maximum'' in (\ref{eqn:para2}).

\section{Reconciliation}
How can we reconcile these various algorithmic speedup metrics?

\subsection{Average Parallelism}
\begin{thm}
Leiserson's $S_\infty$ is identical to the 
\uline{average parallelism}~\cite{ieee,ppa} defined as:
\begin{equation}
A = \dfrac{W}{T} \label{eqn:avgpara}
\end{equation}
where $W$ is the total amount of work (expressed in cpu-seconds, for example) and 
$T$ is the total parallel execution time.
\end{thm}

\begin{prf}
Calculate $W$ using the following procedure:
\begin{enumerate}
\item Start at the top of the DAG
\item At each level where there are nodes, draw a horizontal line through them
\item On each horizontal row calculate the time-node product for each node
\item Sum all the time-node products on each row to get $W$
\end{enumerate}
which can be written symbolically as:
\begin{equation}
W = \sum_{i=1}^{\rm depth} t_i \times n_i 
\end{equation}
For Fig.~\ref{fig:dag} we obtain:
\begin{multline}
W = (1\times1) +
(1\times1) + (1\times1) + (1\times3) + (1\times4) + (1\times4) + \\
(1\times2) + (1\times1) + (1\times1)  \label{eqn:work}
\end{multline}
or $W = 18$.
The value of $T$ can be obtained be simply adding together all the time 
factors in the products of (\ref{eqn:work}), i.e., $T = 9$, since
$t_i = 1$ and there are nine terms. 
Applying (\ref{eqn:avgpara}): $A = 18/9 = 2$, which is identical to (\ref{eqn:para2}).
Thus, $S_\infty \equiv A$. \qed
\end{prf}

\begin{rem}
This is consistent with Leiserson's definition of $S_\infty$ 
as the average amount of work-per-node along the critical path. See Section~\ref{sec:bounds}. 
Other examples of calculating average parallelism are presented in Ref.~\cite{avgpara}. 
\end{rem}

\subsection{Superlinear Performance}
Finally, we can see how Leiserson excludes superlinear performance on the basis of bound (\ref{eqn:linear}).

\begin{thm}
The bound (\ref{eqn:linear}) is equivalent to any computational DAG compressed to depth one.
\end{thm}

\begin{prf}
In Fig.~\ref{fig:dag}, such node compression is equivalent to having  all
18 nodes positioned on the same horizontal row. Since it is not possible  
to squash the DAG any flatter, the best possible speedup 
corresponds to distributing those 18 nodes simultaneously onto $p=18$ processors.
This bound is identical to ideal linear speedup (\ref{eqn:linspeed}), i.e., $S_p = 18$. 
\qed
\end{prf}

\begin{cor}
From (\ref{eqn:efficiency}),
linear speedup corresponds to an efficiency $E_p = 18/18 = 1$.
\end{cor}

Superlinear speedup (\ref{eqn:superlin}) corresponds to an efficiency $E_p
\geq 1$. One way this might be observed is to run the work in
Fig.~\ref{fig:dag} successively on $p = 1,2,3,\ldots$ processors. The
speedup for small-$p$ would be inferior to that for large-$p$, so the
scaling would appear to become better than linear. However, this apparent
improvement is just an artifact of choosing the wrong baseline to establish
linearity in the first place.

\end{document}